\documentclass[preprint,review]{elsarticle}

\journal{}

\usepackage{color}
\usepackage{lscape}
\usepackage{multirow}
\usepackage{amssymb}
\usepackage{amsmath}
\usepackage{amsthm}
\usepackage{url}
\usepackage{graphicx}
\usepackage{hyperref}
\usepackage{rotating}

\usepackage[top=2cm, bottom=2cm, left=2cm, right=2cm]{geometry}
\usepackage{array}
\usepackage{tikz}
\usetikzlibrary{shapes,arrows}
\usetikzlibrary{calc}
\usepackage{booktabs}
\usepackage{tabularx}
\usepackage{longtable} 
\usepackage{graphicx,subfigure}

\newcolumntype{H}{>{\setbox0=\hbox\bgroup}c<{\egroup}@{}}

\begin{document}

\begin{frontmatter}

\title{Code Smell Detection using Multilabel Classification Approach}
\author{Thirupathi Guggulothu}
\ead{thirupathi.gugguloth@gmail.com}
\author{Salman Abdul Moiz\corref{cor1}}
\ead{salman@uohyd.ac.in}

\address{School of Computer and Information Sciences, University of Hyderabad, Hyderabad-500 046, Telangana, India}
\cortext[cor1]{Corresponding author}

\date{}
\begin{abstract}

Code smells are characteristics of the software that indicate a code or design problem which can make software hard to understand, evolve, and maintain. The code smell detection tools proposed in the literature produce different results, as smells are informally defined or are subjective in nature. To address the issue of tool subjectivity, machine learning techniques have been proposed which can learn and distinguish the characteristics of smelly and non-smelly source code elements (classes or methods). However, the existing machine learning techniques can only detect a single type of smell in the code element which does not correspond to a real world scenario. In this paper, we have used multilabel classification methods to detect whether the given code element is affected with multiple smells or not. We have considered two code smell datasets for this work and converted them into a multilabel dataset. In our experimentation, Two multilabel methods performed on the converted dataset which demonstrates good performances in the 10-fold cross-validation, using ten repetitions. 

\end{abstract}


\end{frontmatter}

\section{INTRODUCTION}
\label{sec:1}
Code smell refers to an anomaly in the source code that shows violation of basic design principles such as abstraction, hierarchy, encapsulation, modularity, and modifiability \cite{booch1980object}. Even if the design principles are known to the developers, they are been violated because of inexperience, deadline pressure, and heavy competition in the market. Fowler \textit{et al.} \cite{fowler1999refactoring}, have defined 22 informal code smells. One way to remove them is by using refactoring techniques \cite{opdyke1992refactoring}. Refactoring is a technique that makes better internal structure (design quality) of the code without altering the external behavior of the software. 

\par In the literature, there are several techniques \cite{kessentini2014cooperative} and tools \cite{fontana2012automatic} available to detect different code smells. Each technique and tool produces different results. According to Kessentini \textit{et al.}, the code smell detection techniques can be classified into seven categories (cooperative-based \cite{abdelmoez2014risk}, visualization based \cite{murphy2010interactive}, search-based \cite{palomba2015mining} \cite{liu2013monitor} \cite{palomba2013detecting}, probabilistic \cite{Rao07detectingbad}, metric-based \cite{marinescu2004detection} \cite{moha2010decor} \cite{tsantalis2009identification}, symptoms based \cite{moha2010domain}, and manual techniques \cite{travassos1999detecting} \cite{ciupke1999automatic}) which differs in the underlying algorithm. Bowes \textit{et al.} \cite{bowes2013inconsistent}, compared two smell detection tools on message chaining and showed the disparity of results between them. Because of the differing results, Rasool \textit{et al.} \cite {rasool2015review} have classified, compared and evaluated existing detection tools and techniques to understand the categorization better. The three main reasons for the disparity in the results are: 1) the code smells can be subjectively interpreted by the developers, and hence detected in different ways.(2) Agreement between the detectors is low, i.e., different tools or rules detect a different type of smell for different code elements. 3) The threshold value for identifying the smell can vary for the detectors.

\par To address the above limitations, in particular the subjective nature, Fontana \textit{et al.} \cite{fontana2016comparing} proposed a machine learning (ML) technique to detect four code smells with the help of 32 classification techniques. The authors showed that most of the classifiers achieved more than 95\% performance in terms of accuracy and F-measure. After observing the results, authors have suggested that ML algorithms are most suitable approach for the code smell detection. Di Nucci \textit{et al.} \cite{di2018detecting} addressed some limitations in the Fontana \textit{et al.} \cite{fontana2016comparing}, that the prepared datasets do not represent a real world scenario. That is in the datasets, metric distribution of smelly elements strongly different than the metric distribution of non smelly instances, then any ML technique might easily distinguish the two classes. Where boundary between smelly and non smelly characteristics is not always clear in real case \cite{tufano2017and}, \cite{fontana2016antipattern}. In addition, the authors built four datasets, one for each smell. Each dataset contained code elements (instances) affected by that type of smell or non-smelly components. This makes the datasets unrealistic i.e., a software system usually contains different types of smells and might have made easier for the classifiers to discriminate smelly instances.

\par To over come the above limitations, Di Nucci \textit{et al.} \cite{di2018detecting}, modified the datasets of Fontana \textit{et al.} \cite{fontana2016comparing}, to simulate a more realistic scenario by merging the class and method-level wise datasets. The merged datasets have reduced the metric distribution and contains more than one type of smell instances. The authors experimented the same ML techniques as the Fontana et al., on revised datasets and achieved an average 76\% of accuracy in all models. Their datasets has some instances which are identical but have different class labels called disparity (smelly and non-smelly). In this paper, we addressed the disparity instances and due to this the performances decreased in Di Nucci \textit{et al.} \cite{di2018detecting}. For example, in method level merging, if the long method dataset has an instance which is smelly, and if the same instance is in the feature envy dataset then authors \cite{di2018detecting} replicate that instance in long-method as non-smelly.  This disparity will confuse the ML algorithms. Apart from this issue, the datasets have multiple type code smell instances, but they are not able to detect them.

\par In this work, we removed the disparity instances in the merged method level datasets and experimented tree-based classifiers techniques on them. The results report, an average 95\%- 98\% accuracy for the datasets. There is a drastic change in the performance after removal of disparity. From the datasets of Fontana \textit{et al.}\cite{fontana2016comparing} and Di Nucci \textit{et al.} \cite{di2018detecting}, we have observed that there are 395 common instances in method level. These instances led to an idea to form multilabel dataset. Through this dataset disparity can be eliminated, and more than one smells can be detected for the same instance by using multilabel classification methods. Till now, in the literature \cite{azeem2019machine}, three classification types were used in the code smell detection: 1) binary code smell (presence or absence) 2) based on probability 3) based on severity. 

\par In this paper, we formulate the code smell detection as a multilabel classification (MLC) problem. For this, we considered two method level datasets from Fontana \textit{et al.} \cite{fontana2016comparing} and converted them into multilabel dataset (MLD). We applied, two multilabel classification methods on the dataset. We found that these classification methods achieved good performances (on average 91\%) in the 10-fold cross validation using 10-iterations.
 
\par The structure of the paper is organized as follow; The second section, introduces a work related to detection of code smell using ML techniques; The third section, describes the reference study  of considered datasets; The fourth section, explains the proposed approach; The fifth section, presents experimental setup and results of the proposed study; The sixth section, discusses the proposed study with the previous; The final section, gives conclusion and future directions to this research paper.

\section{Related Work}
\label{Rel}
Over the past fifteen years, researchers presented various tools and techniques for detecting code smells. According to kessentini \textit{et al.} \cite{kessentini2014cooperative} approaches of code smell detection are classified into 7 categories (i.e., cooperative-based approaches, visualization based approaches, machine learning-based approaches, probabilistic approaches, metric-based approaches, symptoms based approaches, and manual approaches). In this section, we consider only machine learning-based approaches for detecting the code smells.

\par Kreimer and Jochen \cite{kreimer2005adaptive}, introduces an adaptive detection to combine known methods for finding design flaws viz., Big Class (Large Class) and Long Method on the basis of metrics with learning decision trees. The analyses were conducted on two software systems known as: IYC system and the WEKA package.

\par Khomh \textit{et al.} \cite{khomh2009bayesian}, propose a Bayesian approach to detect occurrences of the Blob antipattern on open-source programs (GanttProject v1.10.2 and Xerces v2.7.0). \cite{khomh2011bdtex} present BDTEX (Bayesian Detection Expert), a Goal Question Metric approach to build Bayesian Belief Networks from the definitions of antipatterns and validate BDTEX with Blob, Functional Decomposition, and Spaghetti Code antipatterns on two open-source programs.

\par Maneerat \textit{et al.} \cite{maneerat2011bad}, collect datasets from the literature regarding the evaluation of 7 bad-smells, and apply 7 machine learning algorithms for bad-smells prediction, using 27 design model metrics extracted by a tool as independent variables. The author make no explicit reference to the applied datasets.

\par Maiga \textit{et al.} \cite{maiga2012support}, introduce SVMDetect, an approach to detect anti-patterns, based on support vector machines. The subjects of their study are Blob, Functional Decomposition, Spaghetti Code and Swiss Army Knife antipatterns, on three open-source programs: ArgoUML, Azureus, and Xerces. \cite{maiga2012smurf} extend the previous paper by introducing SMURF, which takes into account practitioners’ feedback.

\par Wang \textit{et al.} \cite{wang2012can}, propose an approach that assists in understanding the harmfulness of intended cloning operations using Bayesian Networks and a set of features such as history, code, destination features.

\par Yang \textit{et al.} \cite{yang2015classification}, study the judgment of individual users by applying machine learning algorithms on code clones. \cite{white2016deep}, detected code clone by using deep learning techniques. The authors have sampled 398 files and 480 method levels pairs across 8 real world java software system.

\par Amorim \textit{et al.} \cite{amorim2015experience}, studied the effectivness of the Decision Tree algorithm to recognize code smells. For this, the authors experimented on 4 open source projects and the results were compared with the manual oracle, with existing detection approaches and with other machine learning algorithms. 
\par Fontana \textit{et al.} \cite{fontana2016comparing}, experimented and compared code smell detection through supervised ML algorithms. The author experimented 74 Java systems which are manually validated instances on training dataset and used 16 different classification algorithms. In addition, a boosting techniques is applied on 4 code smells viz., Data Class, Long Method, Feature Envy, God Class.

\par Fontana \textit{et al.} \cite{fontana2017code} , Classified the code smells severity by using a machine learning method. This approach can help software developrs to priortize or rank the classes or methods. Multinomail classifcation and regression were used for code smell severity classification.

\par Di Nucci \textit{et al.} \cite{di2018detecting}, covered some of the limitaions of the Fontana \textit{et al.}\cite{fontana2016comparing}. The authors configured the datasets of Fontana and provided new datasets which are suitable for real case scenario.

\par When observed, the major difference of the previous work with respect to the proposed approach is that the detection of code smells is viewed as multilabel classfication. This paper address some limitations of \cite{di2018detecting} and shown the reason for degraded the results.

\section{Reference Datasets}
In this paper, we consider two method level datasets (long method and feature envy) from Fontana \textit{et al.} \cite{fontana2016comparing}. In existing literature, these datasets are used as a single label methods. In the following subsections, we briefly describe the data preparation methodology of Fontana et al. These datasets are available at \url{http://essere.disco.unimib.it/reverse/MLCSD.html}

\subsection{Systems and Code Smell Selection}
Fontana \textit{et al.} \cite{fontana2016comparing}, have analyzed Qualitus Corpus software systems which are collected from Tempero \textit{et al.} \cite{tempero2010qualitas}. Among 111 systems of the corpus, 74 systems are considered. The remaining 37 systems can not detect code smells as they are not successfully compiled. For the given 74 software systems, the authors have computed 61 class level and 82 method level metrics. These metrics became features for independent variables in the datasets. The two method level code smells used to detect them are long method and feature envy.

\begin{enumerate}
\item \textit{Long Method (LM):} A code smell is said to be long method when it has more number of lines in the code and requires too many parameters. This increases the functional complexity of the method and it will be difficult to understand. 

\item \textit{Feature Envy (FE):} Feature Envy is the method level smell which uses more data from other classes rather than its own class i.e., it accesses more foreign data than the local one.

\end{enumerate}

\subsection{Dataset Preparation}
To establish the dependent variable for code smell prediction models, the authors applied to each code smell a set of automatic detectors shown in Table \ref{t1}. However, code smell detectors cannot usually achieve 100\% recall, meaning that an automatic detection process might not identify actual code smell instances (i.e., false negatives) even in the case that multiple detectors are combined. To cope with false positives and to increase their confidence in validity of the dependent variable, the authors applied a stratified random sampling of the classes/methods of the considered systems: this sampling produced 1,986 instances (826 smelly elements and 1,160 non-smelly ones), which were manually validated by the authors in order to verify the results of the detectors. As a final step, the sampled dataset was normalized for size: the authors randomly removed smelly and non-smelly elements building four disjoint datasets, i.e., one for each code smell type, composed of 140 smelly instances and 280 non-smelly ones (for a total of 420 elements). These datasets represented the training set for the ML techniques.
\begin{table}[h]
\centering
\caption{Automatic Code Smell Detector Tools}
\label{t1}
\begin{tabular}{ll}
\hline
Code Smell   & Detectors                                \\ \hline
Long Method  & PMD \footnote{\url{http://pmd.sourceforge.net/}},  iPlasma(\cite{marinescu2005measurement}), Marinescu(\cite{marinescu2002measurement})           \\
Feature envy & Fluid Tool(\cite{nongpong2012integrating}), iPlasma(\cite{marinescu2005measurement})  \\   
\hline                  
\end{tabular}
\end{table}

\section{Multilabel Classification Approach}
Supervised classification is the task of using algorithms that allow the machine to learn associations between instances and class labels. Supervision comes in the form of previously labeled instances, from which an algorithm builds a model to automatically predict the labels of new instances. In ML, classification problems can be classified into three main categories: Binary (yes or no), MultiClass and Multilabel classification (MLC). In literature \cite{azeem2019machine}, code smell detection were single label (binary) classifiers, used to detect the single type code smell (presence or absence) only. In this work, multilabel classifiers are used to detect the multiple code smells for the same element. 
\par MLC is a way to learn from instances that are associated with a set of labels (predictive classes). That is, for every instance there can be one or more labels associated with them. MLC is frequently used in some application areas like multimedia classification, medical diagnosis, text categorization, and semantic scene classification. Similarly, in our code smell detection domain, instances are code elements and set of labels are code smells, i.e., a code element can contain more than one type of smell which is not addressed by the earlier approaches. The main difference between MLC and existing approaches is that the expected output from the trained models. Existing approaches detected only one smell but, in the proposed one more than one smell can be detected. In the following subsections, we explain the procedure of constructed MLD and methods used for experimentation of multiple label classification.

\subsection{Construction of Multi-label Dataset}
 The considered LM and FE datasets have 420 instances each, which are used to construct multilabel dataset. Following are the steps to create MLD.

\begin{enumerate}
\item Initially, each data set have 420 instances. From those, 395 common instances are added to MLD with their corresponding two class labels.
\item The remaining 25 instances of each single class label dataset are added into MLD by considering the other class label as non smelly.  
\end{enumerate}

\par
\begin{figure}[h]
\centering


  \includegraphics[scale=0.4]{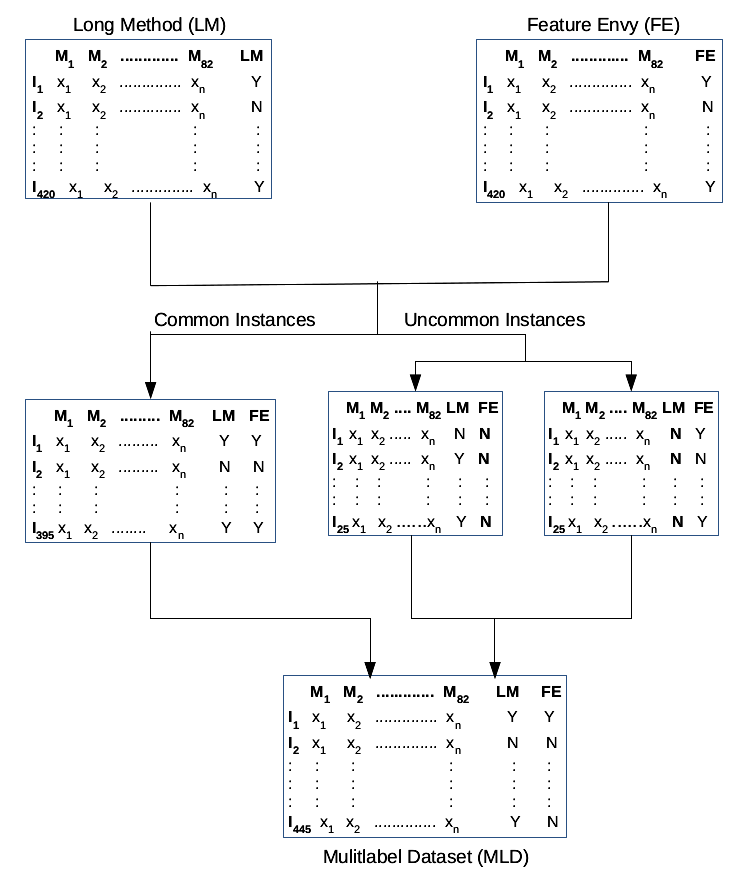}


\caption{Procedure of constructing multilabel dataset}

\label{fig:1}       

\end{figure}

 An overview of the procedure is depicted in Figure \ref{fig:1}. As shown in Figure, the data set contains 82 method metrics namely M1, M2, .. M82 (Independent variables). I1, I2,…... are the instances and the class labels are LM and FE respectively.  

\subsection{Methods of Multilabel Classification}
There are two approaches that are widely used to handle the problems of MLC \cite{tsoumakas2007multi}: problem transformation methods (PTM) and algorithm adoption methods (AAM). In PTM, MLD is transformed to single label problem and are solved by appropriate single label classifiers. In algorithm adaptation, MLD is handled by adapting a single label classifier to solve it. In this paper, we consider only problem transformation method.   
\par There are many methods which fall under PTM category. Among them two methods can be thought of as foundation to many other methods. (1) Binary relevance (BR) method \cite{godbole2004discriminative}: it will convert an MLD to as many binary datasets as the number of different labels that are present. The different dataset predictions from binary classifiers are joined to get the final outcome. (2) Label power set(LP) method \cite{boutell2004learning}: is used to convert MLD to Multi-class dataset based on the label set of each instance as a class identifier. The predicted classes are transformed back to label set using any multi-class classifier. 
\par Several algorithms developed under BR and LP methods. In this paper, there have been two algorithms which covering these methods: Classifier chains (CC) under BR category and LC aka LP category. The reason for choosing these algorithms is that they capture the label dependencies (correlation or co-occurrence) during classification is thus leading to improve the classification performance \cite{guo2011multi}. Usually, the considered code smells co-occur each other \cite{palomba2017investigating}.  
\par After the transformation, we used top 5 tree based (single label) classifiers for the predictions of multilabel methods (CC, LC). In the literature \cite{azeem2019machine}, previous studies shown that, these classifiers achieved high performance in the code smell classification.     

We have identified set of specific research questions which guides to classify the code smells using multilabel approach:
\par \textbf{RQ1:} How many disparity instances are existing in the configured datasets of the concerned code smells in the \cite{di2018detecting}. 
\par \textbf{RQ2:} What would be the performance improvement after removing the disparity instances?
\par \textbf{RQ3:} What would be the performance when constructed the dataset by using multilabel instead of merging?

\section{Experimental Results}
\subsection{Experimentation Setup}
In the following, report the MLC methods with a short description and MEkA \cite{read2016meka} tool provides the implementation of the selected methods.
\begin{itemize}

\item Classifier Chains (CC) \cite{read2011classifier}: The algorithm tries to enhance BR by considering the label correlation. To predict the new labels, train ’Q’ classifiers which are connected to one another in such a way that the prediction of each classifier is being added to the dataset as new feature.

\item LC aka LP (Label Powerset) Method \cite{boutell2004learning}: Treats each label combination as a single class in a multi-class learning scheme. The set of possible values of each class is the powerset of labels. 
\end{itemize}
\par To test the performance of the different code smell prediction models built, we apply 10-fold cross validation and run them up to 10 times to cope with randomness \cite{hall2011developing}. Next, we evaluate the classification performance.
\par The evaluation metric of MLC is different from that of single label classification, since for each instance there are multiple labels which may be classified partly correctly or partly incorrectly. MLC evaluation metrics are classified into two groups: (1) Example based metrics (2) Label based metrics. In example based metrics one each instance metric is calculated and then average of those metrics gives the final outcome. Label based metrics are computed for each label instead of each instance. In this work, we have considered example based measures. Label based measures would fail to directly address the correlations among different classes \cite{sorower2010literature}. Below, equations \ref{eq:1}, \ref{eq:2}, and \ref{eq:3} are used to measure the performances of MLC methods which belongs to the example-based category. In this, D denotes number of instances, L represents number of labels, $Y_i$ is the predicted labels for instance i, and $Z_i$ denotes true labels for instance i. Detailed discussion of all other measures are defined in \cite{sorower2010literature}.

\begin{itemize}
\item \textbf{Accuracy:} The proportion of correctly predicted labels with respect to the number of labels for each instance. 
	\begin{equation} \label{eq:1} 
		 Accuracy = \frac{1}{|D|} \sum_{i=1}^{|D|} \frac{|Y_{i} \cap Z_{i}|} {|Y_{i} \cup  Z_{i}|} 
	\end{equation}
\item \textbf{Hamming Loss:} The prediction error (an incorrect label is predicted) and the missing error (a relevant label not predicted), normalized over total number of classes and total number of examples.	
	\begin{equation} \label{eq:2} 
		Hamming Loss = \frac{1}{|D|} \sum_{i=1}^{|D|} \frac{|Y_{i} \Delta Z_{i}|} {|L|}
	\end{equation}
\item \textbf{Exact match Ratio:} The predicted label set is identical to the actual label set. It is a most strict evaluation metric. 
	\begin{equation} \label{eq:3} 
		 Exact match Ratio = \frac{1}{|D|} \sum_{i=1}^{|D|} I(Y_{i}=Z_{i})
	\end{equation}
\end{itemize}

\subsection{Results}
\subsubsection{Datasets}
To answer the \textbf{RQ1}, we have considered the configured datasets of \cite{di2018detecting}. The author merged the FE dataset into LM dataset and vice versa. The merged datasets are listed in Table \ref{t2}. In a table, each dataset has 840 instances, among them 140 instances affected (smelly) and 700 are non-smelly. While merging FE into LM, there are 395 common instances among which 132 are smelly instances in LM dataset. In the same way, when LM is merged with FE, there are 125 smelly instances in FE dataset. These 132 and 125 instances are suffered from disparity i.e., same instance is having two class label (smelly and non-smelly). Due to the disparity instances \cite{di2018detecting}, authors achieved less performances in the ML classification techniques. The merged datasets are available at \url{https://figshare.com/articles/Detecting_Code_Smells_using_Machine_Learning_Techniques_Are_We_There_Yet_/5786631}
\begin{table}[h]
\centering
\caption{Configured Datasets}
\label{t2}
\begin{tabular}{ccccc}
\hline
\multirow{3}{*}{\begin{tabular}[c]{@{}c@{}}Number of \\ Instances\end{tabular}} & \multicolumn{4}{c}{Method Level Merged Datasets}                                                                                                                                                                                                             \\ \cline{2-5} 
                                                                                & \multicolumn{2}{c}{Long Method}                                                                                               & \multicolumn{2}{c}{Feature Envy}                                                                                             \\ \cline{2-5} 
                                                                                & \begin{tabular}[c]{@{}c@{}}Smelly\\  Instances\end{tabular} & \begin{tabular}[c]{@{}c@{}}Non Smelly\\  Instances\end{tabular} & \begin{tabular}[c]{@{}c@{}}Smelly\\  Instance\end{tabular} & \begin{tabular}[c]{@{}c@{}}Non Smelly \\ Instances\end{tabular} \\ \hline
840                                                                             & 140                                                         & 700                                                             & 140                                                        & 700  \\ \hline                                                          
\end{tabular}
\end{table}
  
\par
\begin{figure}[h]
\centering


  \includegraphics[scale=0.45]{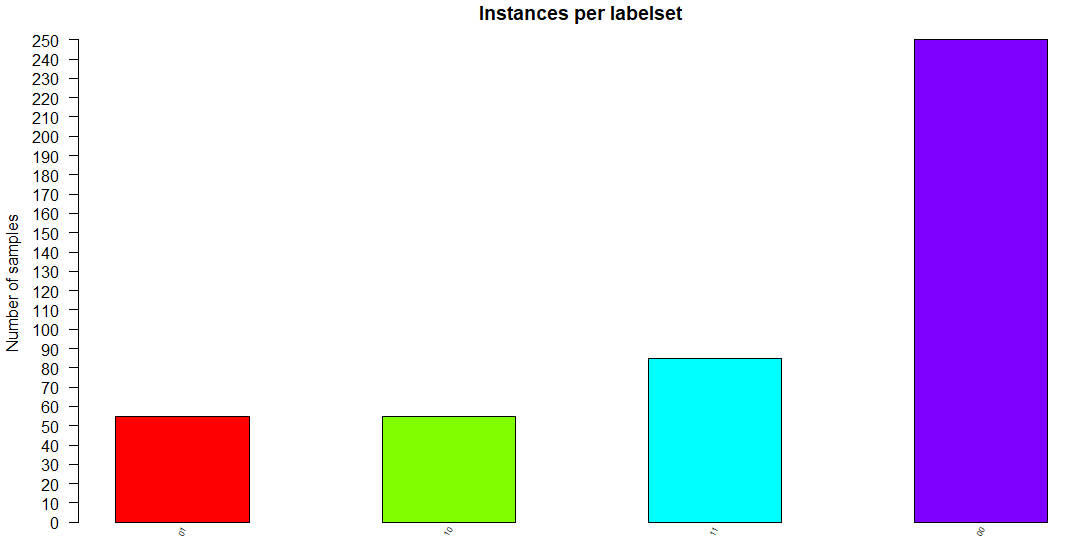}


\caption{Multilabel dataset}

\label{fig:2}       

\end{figure}

 In this paper, MLD is created by considering 395 common and 50 uncommon (25 each) instances of LM and FE merged; there are 445 instances. Table \ref{t3} shows the percentage and number of instances affected in the MLD. Out of 445, 85 instances are affected by both the smells. When concerned individually there are 140 instances affected by LM and FE. The grahphical representation of MLD is shown in Figure \ref{fig:2}.

\begin{table}[h]
\centering
\caption{Number of instances affected in multilabel dataset}
\label{t3}
\begin{tabular}{cccc}
\hline
\begin{tabular}[c]{@{}c@{}}Long Method\\ Affected\end{tabular} & \begin{tabular}[c]{@{}c@{}}Feature Envy\\ Affected\end{tabular} & \begin{tabular}[c]{@{}c@{}}Number of \\ Instances Affected\end{tabular} & \% of Affected \\ \hline
Yes                                                            & Yes                                                             & 85                                                                      & 19.1\%         \\
Yes                                                            & No                                                              & 55                                                                      & 12.3\%         \\
No                                                             & Yes                                                             & 55                                                                      & 12.3\%         \\
No                                                             & No                                                              & 250                                                                     & 56.17\%         \\
\hline
\end{tabular}
\end{table} 

\subsubsection{Multilabel Dataset Statistics}
Table \ref{t4} lists the basic measures of multi-label training dataset characteristics. Some of the basic measures in single label dataset are attributes, instances, and labels. In addition to it there are other measures added to multilabel dataset \cite{tsoumakas2007multi}. In the table, cardinality indicates the average number of active labels per instance. Dividing this measure by number of labels in dataset, results in a dimensionless measure known as density. The two labels will have four label combinations (label sets) in our dataset. The mean imbalance ratio (mean IR) gives the information about, whether the dataset is imbalanced or not. As a general rule, \cite{charte2015addressing} any MLD with a MeanIR value higher than 1.5 should be considered as imbalanced. With this, the prepared multilabel dataset is well balanced because of the MeanIR value in our case is 1.0 which is less than the 1.5. 
\begin{table}[h]
\centering
\caption{ Statistics of Multilabel Dataset }
\label{t4}
\begin{tabular}{lllllll}
\hline
\begin{tabular}[c]{@{}l@{}}Number of\\ Instances\end{tabular} & \begin{tabular}[c]{@{}l@{}}Number of\\ Featrues\end{tabular} & \begin{tabular}[c]{@{}l@{}}Number of\\ Labels\end{tabular} & \begin{tabular}[c]{@{}l@{}}Number of\\ Label Sets\end{tabular} & Cardinality & Density & MeanIR \\ \hline
445                                                           & 82                                                           & 2                                                          & 4                                                              & 0.629       & 0.314   & 1.0 \\ \hline
\end{tabular}
\end{table}

\subsubsection{ Performance Improvements in Existing Datasets}
To answer \textbf{RQ2}, We have removed 132, and 125 disparity instances of LM and FE merged datasets respectively. Now, the LM dataset has 708 instances among them 140 are positive (Smelly), and 568 are negative (non-smelly). In FE dataset has 715 instances among them 140 are positive, and 575 are negative. Then, we used single label ML techniques (tree based classifiers) on those datasets. Now, the performance got drastically improved on both the datasets which are shown in Tables \ref{t5} and \ref{t6}. Earlier the performance on long method and feature envy datasets were an average 73\% and 75\% using tree based classifier. After removal of disparity instances in both the datasets, now we got an average 95\%, 98\%.  With this evidence, due to disparity, Di Nucci \textit{et al.}\cite{di2018detecting} got less performance on the concerned code smell datasets. 
\begin{table}[h]
\centering
\caption{Long Method Results}
\label{t5}
\begin{tabular}{cccc}
\hline
Classifier      & Accuracy & F-Measure & ROC Area \\ \hline
B-Random Forest & 95.9\%   & 96.0\%    & 97.6\%   \\
Random Forest   & \textbf{95.9\%}   & \textbf{96.0\%}    & \textbf{97.7\%}   \\
B-J48 UnPruned  & 95.4\%   & 95.5\%    & 97.1\%   \\
B-J48 Pruned    & 94.7\%   & 94.8\%    & 97.7\%   \\
J48 Unpruned    & 93.5\%   & 93.5\%    & 91.9\%  \\ \hline
\end{tabular}
\end{table}

\begin{table}[h]
\centering
\caption{Feature Envy Results}
\label{t6}
\begin{tabular}{cccc}
\hline
Classifier      & Accuracy & F-Measure & ROC Area \\ \hline
B-Random Forest & 98.0\%   & 98.0\%    & 99.9\%   \\
Random Forest   & 98.1\%   & 98.2\%    & 99.9\%   \\
B-J48 UnPruned  & 99.0\%   & 99.0\%    & 98.7\%   \\
B-J48 Pruned    & \textbf{99.1\%}   & \textbf{99.2\%}    & \textbf{99.3\%}   \\
J48 Unpruned    & 98.1\%   & 98.2\%    & 98.0\%  \\ \hline
\end{tabular}
\end{table}

\subsubsection{ Performances of Multilabel classfication}
To answer the \textbf{RQ3}: Two problem transformation methods (CC, LC) are used to transform multi-label training dataset into a set of binary or multi-class datasets. Then, we have used top 5 tree-based classification techniques on the transformed dataset. The performances of those techniques are shown in the tables respectively \ref{t7} and \ref{t8}. To evaluate the techniques, we have run them for 10 iterations using 10 fold cross-validation. We measured average accuracy, hamming loss, and an exact match of those 100 iterations.
\par From the tables \ref{t7}, \ref{t8} reports that all top 5 classifiers performing well under the CC, LC methods. The best results report 89.6\%-93.6\% accuracy for CC and 89\%-93.5\% for LC method with low hamming loss $<$ 0.079 in most cases. In both the tables, it is shown that random forest classifier is giving the best performance based on all three measures. As a method wise, CC method performing slight over the LC method. In addition to these results, we also listed other metrics (label-based) of CC and LC methods which are reported in Appendix table \ref{t9} and \ref{t10}.    

\begin{table}[h]
\centering
\caption{Results of CC method using top 5 single label classifers.}
\label{t7}
\begin{tabular}{cccc}
\hline
\multicolumn{4}{c}{CC (10-Fold Cross Validation Run for 10 Iterations)}                                                                                                                                                                                                   \\ \hline
\multirow{2}{*}{\begin{tabular}[c]{@{}c@{}}Single Label\\ Classifier\end{tabular}} & \multicolumn{3}{c}{Example Based Metrics}                                                                                                                                            \\ \cline{2-4} 
                                                                                   & \begin{tabular}[c]{@{}c@{}}Accuracy\\ (Jaccard Index)\end{tabular} & \begin{tabular}[c]{@{}c@{}}Hamming\\ Loss\end{tabular} & \begin{tabular}[c]{@{}c@{}}Exact\\ Match\end{tabular} \\ \hline
J48 Pruned                                                                         & 89.6\%                                                             & 0.078                                                  & 85.4\%                                                \\
Random Forest                                                                      & \textbf{93.6\%}                                                    & \textbf{0.047}                                         & \textbf{91.1\%}                                       \\
B-J48 pruned                                                                       & 92.2\%                                                             & 0.056                                                  & 89.4\%                                                \\
B-J48 UnPruned                                                                     & 91.1\%                                                             & 0.064                                                  & 87.9\%                                                \\
B-Random Forest                                                                    & 92.8\%                                                             & 0.053                                                  & 89.9\%       \\ \hline                                        
\end{tabular}
\end{table}

\begin{table}[h]
\centering
\caption{Results of LC method using top 5 single label classifers.}
\label{t8}
\begin{tabular}{cccc}
\hline
\multicolumn{4}{c}{LC (10-Fold Cross Validation Run for 10 Iterations)}                                                                                                                                                                                                   \\ \hline
\multirow{2}{*}{\begin{tabular}[c]{@{}c@{}}Single Label\\ Classifier\end{tabular}} & \multicolumn{3}{c}{Example Based Metrics}                                                                                                                                            \\ \cline{2-4} 
                                                                                   & \begin{tabular}[c]{@{}c@{}}Accuracy\\ (Jaccard Index)\end{tabular} & \begin{tabular}[c]{@{}c@{}}Hamming\\ Loss\end{tabular} & \begin{tabular}[c]{@{}c@{}}Exact\\ Match\end{tabular} \\ \hline
J48 Pruned                                                                         & 89.0\%                                                             & 0.075                                                  & 85.2\%                                                \\
Random Forest                                                                      & 93.3\%                                                             & 0.053                                                  & 90.1\%                                                \\
B-J48 pruned                                                                       & 90.0\%                                                             & 0.069                                                  & 87.0\%                                                \\
B-J48 UnPruned                                                                     & 90.7\%                                                             & 0.063                                                  & 87.9\%                                                \\
B-Random Forest                                                                    & \textbf{93.5\%}                                                    & \textbf{0.049}                                         & \textbf{90.6\%}   \\ \hline                                   
\end{tabular}
\end{table}

\par The LC method aka LP is used to convert MLD to Multi-class dataset based on the label set of each instance as a class identifier. That is, in this work, a multiclass can contains four class (00,01,10,11) values, 00 means not affected by both smells, 01 means affected by feature envy, 10 means affected by long method, and 11 means affected by both the smells. Table \ref{t8}, also said the results of Multiclass classification.

\subsection{Discussion}
In this section, we discuss how the existing studies differ from the proposed study. Then, we give how our proposed approach is much more useful in a real-world scenario.
\par The study \cite{di2018detecting}, replicated and modified the datasets of \cite{fontana2016comparing} by merging the instances of other code smell datasets to i)reduce the difference in the metric distribution ii) have the different type of smells in the same dataset so that can model a more realistic scenario.  

\par In this paper, we identified the disparity instances in the merged datasets and removed them by manual process. After that, we used the same tree-based classifiers as in the \cite{di2018detecting} on the removal disparity instances datasets and achieved 95\% and 98\% accuracy in LM and FE respectively. This disparity will lead to forming the idea of multilable dataset.  

\par In a real-world scenario, a code element can contain more than one design problems (code smells) and our MLD constructed accordingly. The MLD also maintain similar characteristics as in the modified datasets of \cite{di2018detecting}, like metric distribution and have different types of smells. Then, two MLC methods used on the MLD. In the existing study, the performances were an average 76\% accuracy and detected only one type of smell. But, in the proposed study we detected two smells in the same instance and obtained 91\% of accuracy. Our findings have important implications for further research community to 1) analyze the detected code smells after the detection so that which smell is first to refactor to reduce developer effort because different smell orders require different effort 2) Identify (or prioritize) the critical code elements for refactoring based on the number of code smells it detected. That is, if an element can be affected by more design problems then this element given has the highest priority for refactoring.



\section{Conclusion and Future Directions}
In this work, we detected two method level code smells using a multilabel classification approach. Existing studies used to detect a single type code smell but, in the proposed study, we detected two code smells whether they exist in the same method or not. For this work, we considered two method datasets which are constructed by single type detectors. These datasets have 395 common instances thus leads to form the disparity while merging process in the existing study. Due to this, the performances were less in their study. In this paper, these common instances are led to construct the MLD and also to avoid the disparity. We experimented, two multilabel classification methods(CC, LC) on the MLD. The CC method has given best performance than LC based on all three measures. The performance of the proposed study is much better than the existing study. In the existing study, the performance of all models got an average 73\% accuracy, whereas in proposed study we got an average 91\%.   

\par Proposed approach detected only two smells, and it is not limited. In the future, we want to detect other method level code smells also. In addition, the importance of multilabel classification for code smell can identify the critical code elements (method or class) which are urgent need of refactoring. That is, we are classifying the critical element by using multilabel classification based on the number of code smell detected by the element in the dataset. For example, if there are two code smells in the same method, then this method is suffering from more design problems (critical) associated to those code smells rather than single code smell.  
\par The removal of disparity instances datasets are avaliable for download at \url{https://github.com/thiru578/Datasets-LM-FE}.
\par The multilabel dataset available for download at \url{https://github.com/thiru578/Multilabel-Dataset}.


\bibliographystyle{elsarticle-num}
\bibliography{thiru}

\begin{thebibliography}{10}
\expandafter\ifx\csname url\endcsname\relax
  \def\url#1{\texttt{#1}}\fi
\expandafter\ifx\csname urlprefix\endcsname\relax\def\urlprefix{URL }\fi
\expandafter\ifx\csname href\endcsname\relax
  \def\href#1#2{#2} \def\path#1{#1}\fi

\bibitem{booch1980object}
G.~Booch, Object-oriented analysis and design, Addison-Wesley, 1980.

\bibitem{fowler1999refactoring}
M.~Fowler, K.~Beck, J.~Brant, W.~Opdyke, D.~Roberts, Refactoring: Improving the
  design of existing programs (1999).

\bibitem{opdyke1992refactoring}
W.~F. Opdyke, Refactoring: A program restructuring aid in designing
  object-oriented application frameworks, Ph.D. thesis, PhD thesis, University
  of Illinois at Urbana-Champaign (1992).

\bibitem{kessentini2014cooperative}
W.~Kessentini, M.~Kessentini, H.~Sahraoui, S.~Bechikh, A.~Ouni, A cooperative
  parallel search-based software engineering approach for code-smells
  detection, IEEE Transactions on Software Engineering 40~(9) (2014) 841--861.

\bibitem{fontana2012automatic}
F.~A. Fontana, P.~Braione, M.~Zanoni, Automatic detection of bad smells in
  code: An experimental assessment., Journal of Object Technology 11~(2) (2012)
  5--1.

\bibitem{abdelmoez2014risk}
W.~Abdelmoez, E.~Kosba, A.~F. Iesa, Risk-based code smells detection tool, in:
  The International Conference on Computing Technology and Information
  Management (ICCTIM2014), The Society of Digital Information and Wireless
  Communication, 2014, pp. 148--159.

\bibitem{murphy2010interactive}
E.~Murphy-Hill, A.~P. Black, An interactive ambient visualization for code
  smells, in: Proceedings of the 5th international symposium on Software
  visualization, ACM, 2010, pp. 5--14.

\bibitem{palomba2015mining}
F.~Palomba, G.~Bavota, M.~Di~Penta, R.~Oliveto, D.~Poshyvanyk, A.~De~Lucia,
  Mining version histories for detecting code smells, IEEE Transactions on
  Software Engineering 41~(5) (2015) 462--489.

\bibitem{liu2013monitor}
H.~Liu, X.~Guo, W.~Shao, Monitor-based instant software refactoring, IEEE
  Transactions on Software Engineering (2013) 1.

\bibitem{palomba2013detecting}
F.~Palomba, G.~Bavota, M.~Di~Penta, R.~Oliveto, A.~De~Lucia, D.~Poshyvanyk,
  Detecting bad smells in source code using change history information, in:
  Proceedings of the 28th IEEE/ACM International Conference on Automated
  Software Engineering, IEEE Press, 2013, pp. 268--278.

\bibitem{Rao07detectingbad}
A.~A. Rao, K.~N. Reddy, Detecting bad smells in object oriented design using
  design change propagation probability matrix 1 (2007).

\bibitem{marinescu2004detection}
R.~Marinescu, Detection strategies: Metrics-based rules for detecting design
  flaws, in: Software Maintenance, 2004. Proceedings. 20th IEEE International
  Conference on, IEEE, 2004, pp. 350--359.

\bibitem{moha2010decor}
N.~Moha, Y.-G. Gueheneuc, A.-F. Duchien, et~al., Decor: A method for the
  specification and detection of code and design smells, IEEE Transactions on
  Software Engineering (TSE) 36~(1) (2010) 20--36.

\bibitem{tsantalis2009identification}
N.~Tsantalis, A.~Chatzigeorgiou, Identification of move method refactoring
  opportunities, IEEE Transactions on Software Engineering 35~(3) (2009)
  347--367.

\bibitem{moha2010domain}
N.~Moha, Y.-G. Gu{\'e}h{\'e}neuc, A.-F. Le~Meur, L.~Duchien, A.~Tiberghien,
  From a domain analysis to the specification and detection of code and design
  smells, Formal Aspects of Computing 22~(3-4) (2010) 345--361.

\bibitem{travassos1999detecting}
G.~Travassos, F.~Shull, M.~Fredericks, V.~R. Basili, Detecting defects in
  object-oriented designs: using reading techniques to increase software
  quality, in: ACM Sigplan Notices, Vol.~34, ACM, 1999, pp. 47--56.

\bibitem{ciupke1999automatic}
O.~Ciupke, Automatic detection of design problems in object-oriented
  reengineering, in: Technology of Object-Oriented Languages and Systems, 1999.
  TOOLS 30 Proceedings, IEEE, 1999, pp. 18--32.

\bibitem{bowes2013inconsistent}
D.~Bowes, D.~Randall, T.~Hall, The inconsistent measurement of message chains,
  in: Emerging Trends in Software Metrics (WETSoM), 2013 4th International
  Workshop on, IEEE, 2013, pp. 62--68.

\bibitem{rasool2015review}
G.~Rasool, Z.~Arshad, A review of code smell mining techniques, Journal of
  Software: Evolution and Process 27~(11) (2015) 867--895.

\bibitem{fontana2016comparing}
F.~A. Fontana, M.~V. M{\"a}ntyl{\"a}, M.~Zanoni, A.~Marino, Comparing and
  experimenting machine learning techniques for code smell detection, Empirical
  Software Engineering 21~(3) (2016) 1143--1191.

\bibitem{di2018detecting}
D.~Di~Nucci, F.~Palomba, D.~A. Tamburri, A.~Serebrenik, A.~De~Lucia, Detecting
  code smells using machine learning techniques: are we there yet?, in: 2018
  IEEE 25th International Conference on Software Analysis, Evolution and
  Reengineering (SANER), IEEE, 2018, pp. 612--621.

\bibitem{tufano2017and}
M.~Tufano, F.~Palomba, G.~Bavota, R.~Oliveto, M.~Di~Penta, A.~De~Lucia,
  D.~Poshyvanyk, When and why your code starts to smell bad (and whether the
  smells go away), IEEE Transactions on Software Engineering 43~(11) (2017)
  1063--1088.

\bibitem{fontana2016antipattern}
F.~A. Fontana, J.~Dietrich, B.~Walter, A.~Yamashita, M.~Zanoni, Antipattern and
  code smell false positives: Preliminary conceptualization and classification,
  in: Software Analysis, Evolution, and Reengineering (SANER), 2016 IEEE 23rd
  International Conference on, Vol.~1, IEEE, 2016, pp. 609--613.

\bibitem{azeem2019machine}
M.~I. Azeem, F.~Palomba, L.~Shi, Q.~Wang, Machine learning techniques for code
  smell detection: A systematic literature review and meta-analysis,
  Information and Software Technology.

\bibitem{kreimer2005adaptive}
J.~Kreimer, Adaptive detection of design flaws, Electronic Notes in Theoretical
  Computer Science 141~(4) (2005) 117--136.

\bibitem{khomh2009bayesian}
F.~Khomh, S.~Vaucher, Y.-G. Gu{\'e}h{\'e}neuc, H.~Sahraoui, A bayesian approach
  for the detection of code and design smells, in: Quality Software, 2009.
  QSIC'09. 9th International Conference on, IEEE, 2009, pp. 305--314.

\bibitem{khomh2011bdtex}
F.~Khomh, S.~Vaucher, Y.-G. Gu{\'e}h{\'e}neuc, H.~Sahraoui, Bdtex: A gqm-based
  bayesian approach for the detection of antipatterns, Journal of Systems and
  Software 84~(4) (2011) 559--572.

\bibitem{maneerat2011bad}
N.~Maneerat, P.~Muenchaisri, Bad-smell prediction from software design model
  using machine learning techniques, in: Computer Science and Software
  Engineering (JCSSE), 2011 Eighth International Joint Conference on, IEEE,
  2011, pp. 331--336.

\bibitem{maiga2012support}
A.~Maiga, N.~Ali, N.~Bhattacharya, A.~Saban{\'e}, Y.-G. Gu{\'e}h{\'e}neuc,
  G.~Antoniol, E.~A{\"\i}meur, Support vector machines for anti-pattern
  detection, in: Automated Software Engineering (ASE), 2012 Proceedings of the
  27th IEEE/ACM International Conference on, IEEE, 2012, pp. 278--281.

\bibitem{maiga2012smurf}
A.~Maiga, N.~Ali, N.~Bhattacharya, A.~Sabane, Y.-G. Gueheneuc, E.~Aimeur,
  Smurf: A svm-based incremental anti-pattern detection approach, in: Reverse
  engineering (WCRE), 2012 19th working conference on, IEEE, 2012, pp.
  466--475.

\bibitem{wang2012can}
X.~Wang, Y.~Dang, L.~Zhang, D.~Zhang, E.~Lan, H.~Mei, Can i clone this piece of
  code here?, in: Proceedings of the 27th IEEE/ACM International Conference on
  Automated Software Engineering, ACM, 2012, pp. 170--179.

\bibitem{yang2015classification}
J.~Yang, K.~Hotta, Y.~Higo, H.~Igaki, S.~Kusumoto, Classification model for
  code clones based on machine learning, Empirical Software Engineering 20~(4)
  (2015) 1095--1125.

\bibitem{white2016deep}
M.~White, M.~Tufano, C.~Vendome, D.~Poshyvanyk, Deep learning code fragments
  for code clone detection, in: Proceedings of the 31st IEEE/ACM International
  Conference on Automated Software Engineering, ACM, 2016, pp. 87--98.

\bibitem{amorim2015experience}
L.~Amorim, E.~Costa, N.~Antunes, B.~Fonseca, M.~Ribeiro, Experience report:
  Evaluating the effectiveness of decision trees for detecting code smells, in:
  Software Reliability Engineering (ISSRE), 2015 IEEE 26th International
  Symposium on, IEEE, 2015, pp. 261--269.

\bibitem{fontana2017code}
F.~A. Fontana, M.~Zanoni, Code smell severity classification using machine
  learning techniques, Knowledge-Based Systems 128 (2017) 43--58.

\bibitem{tempero2010qualitas}
E.~Tempero, C.~Anslow, J.~Dietrich, T.~Han, J.~Li, M.~Lumpe, H.~Melton,
  J.~Noble, The qualitas corpus: A curated collection of java code for
  empirical studies, in: Software Engineering Conference (APSEC), 2010 17th
  Asia Pacific, IEEE, 2010, pp. 336--345.

\bibitem{marinescu2005measurement}
R.~Marinescu, Measurement and quality in object-oriented design, in: Software
  Maintenance, 2005. ICSM'05. Proceedings of the 21st IEEE International
  Conference on, IEEE, 2005, pp. 701--704.

\bibitem{marinescu2002measurement}
R.~Marinescu, Measurement and quality in objectoriented design.

\bibitem{nongpong2012integrating}
K.~Nongpong, Integrating" code smells" detection with refactoring tool support.

\bibitem{tsoumakas2007multi}
G.~Tsoumakas, I.~Katakis, Multi-label classification: An overview,
  International Journal of Data Warehousing and Mining (IJDWM) 3~(3) (2007)
  1--13.

\bibitem{godbole2004discriminative}
S.~Godbole, S.~Sarawagi, Discriminative methods for multi-labeled
  classification, in: Pacific-Asia conference on knowledge discovery and data
  mining, Springer, 2004, pp. 22--30.

\bibitem{boutell2004learning}
M.~R. Boutell, J.~Luo, X.~Shen, C.~M. Brown, Learning multi-label scene
  classification, Pattern recognition 37~(9) (2004) 1757--1771.

\bibitem{guo2011multi}
Y.~Guo, S.~Gu, Multi-label classification using conditional dependency
  networks, in: IJCAI Proceedings-International Joint Conference on Artificial
  Intelligence, Vol.~22, 2011, p. 1300.

\bibitem{palomba2017investigating}
F.~Palomba, R.~Oliveto, A.~De~Lucia, Investigating code smell co-occurrences
  using association rule learning: A replicated study, in: Machine Learning
  Techniques for Software Quality Evaluation (MaLTeSQuE), IEEE Workshop on,
  IEEE, 2017, pp. 8--13.

\bibitem{read2016meka}
J.~Read, P.~Reutemann, B.~Pfahringer, G.~Holmes, Meka: a
  multi-label/multi-target extension to weka, The Journal of Machine Learning
  Research 17~(1) (2016) 667--671.

\bibitem{read2011classifier}
J.~Read, B.~Pfahringer, G.~Holmes, E.~Frank, Classifier chains for multi-label
  classification, Machine learning 85~(3) (2011) 333.

\bibitem{hall2011developing}
T.~Hall, S.~Beecham, D.~Bowes, D.~Gray, S.~Counsell, Developing
  fault-prediction models: What the research can show industry, IEEE software
  28~(6) (2011) 96--99.

\bibitem{sorower2010literature}
M.~S. Sorower, A literature survey on algorithms for multi-label learning,
  Oregon State University, Corvallis 18.

\bibitem{charte2015addressing}
F.~Charte, A.~J. Rivera, M.~J. del Jesus, F.~Herrera, Addressing imbalance in
  multilabel classification: Measures and random resampling algorithms,
  Neurocomputing 163 (2015) 3--16.

\end{thebibliography}

\newpage
\section*{Appendix}

\begin{table}[h]
\centering
\caption{Results of CC method (example based) using top 5 single label classifers.}
\label{t9}
\begin{tabular}{cccclll}
\hline
\multicolumn{7}{c}{CC (10-Fold Cross Validation Run for 10 Iterations)}                                                                                                                                                                                                                                                                                                                                        \\ \hline
\multirow{3}{*}{\begin{tabular}[c]{@{}c@{}}Single Label\\ Classifier\end{tabular}} & \multicolumn{6}{c}{Label Based Metrics}                                                                                                                                                                                                                                                                                   \\ \cline{2-7} 
                                                                                   & \multicolumn{3}{c}{Micro Averaging}                                                                                           & \multicolumn{3}{c}{Macro Averaging}                                                                                                                                                       \\ \cline{2-7} 
                                                                                   & \begin{tabular}[c]{@{}c@{}}Micro\\ Precision\end{tabular} & \begin{tabular}[c]{@{}c@{}}Micro\\ Recall\end{tabular} & F1-Micro & \multicolumn{1}{c}{\begin{tabular}[c]{@{}c@{}}Macro\\ Precision\end{tabular}} & \multicolumn{1}{c}{\begin{tabular}[c]{@{}c@{}}Macro\\ Recall\end{tabular}} & \multicolumn{1}{c}{F1-Macro} \\ \hline
J48 Pruned                                                                         & 89.0\%                                                    & 95.0\%                                                 & 91.9\%   & 89.1\%                                                                        & 95.0\%                                                                     & 91.9\%                       \\
Random Forest                                                                      & 89.9\%                                                    & 95.7\%                                                 & 92.8\%   & 90.2\%                                                                        & 95.7\%                                                                     & 95.4\%                       \\
B-J48 pruned                                                                       & 89.7\%                                                    & 92.9\%                                                 & 91.2\%   & 89.8\%                                                                        & 92.9\%                                                                     & 913\%                        \\
B-J48 UnPruned                                                                     & 88.6\%                                                    & 91.4\%                                                 & 90.0\%   & 88.7\%                                                                        & 91.4\%                                                                     & 90.0\%                       \\
B-Random Forest                                                                    & 89.0\%                                                    & 95.0\%                                                 & 91.9\%   & 89.1\%                                                                        & 95.0\%                                                                     & 94.8\%                      
\\ \hline
\end{tabular}
\end{table}

\begin{table}[h]
\centering
\caption{Results of LC method (example based) using top 5 single label classifers.}
\label{t10}
\begin{tabular}{cccclll}
\hline
\multicolumn{7}{c}{CC (10-Fold Cross Validation Run for 10 Iterations)}                                                                                                                                                                                                                                                                                                                                        \\ \hline
\multirow{3}{*}{\begin{tabular}[c]{@{}c@{}}Single Label\\ Classifier\end{tabular}} & \multicolumn{6}{c}{Label Based Metrics}                                                                                                                                                                                                                                                                                   \\ \cline{2-7} 
                                                                                   & \multicolumn{3}{c}{Micro Averaging}                                                                                           & \multicolumn{3}{c}{Macro Averaging}                                                                                                                                                       \\ \cline{2-7} 
                                                                                   & \begin{tabular}[c]{@{}c@{}}Micro\\ Precision\end{tabular} & \begin{tabular}[c]{@{}c@{}}Micro\\ Recall\end{tabular} & F1-Micro & \multicolumn{1}{c}{\begin{tabular}[c]{@{}c@{}}Macro\\ Precision\end{tabular}} & \multicolumn{1}{c}{\begin{tabular}[c]{@{}c@{}}Macro\\ Recall\end{tabular}} & \multicolumn{1}{c}{F1-Macro} \\ \hline
J48 Pruned                                                                         & 88.2\%                                                    & 87.9\%                                                 & 88.0\%   & 88.1\%                                                                        & 87.9\%                                                                     & 88.0\%                       \\
Random Forest                                                                      & 88.4\%                                                    & 95.7\%                                                 & 91.9\%   & 88.6\%                                                                        & 95.7\%                                                                     & 92.0\%                       \\
B-J48 pruned                                                                       & 88.7\%                                                    & 89.6\%                                                 & 89.2\%   & 88.8\%                                                                        & 89.6\%                                                                     & 89.2\%                       \\
B-J48 UnPruned                                                                     & 86.2\%                                                    & 91.1\%                                                 & 90.1\%   & 89.2\%                                                                        & 91.1\%                                                                     & 90.1\%                       \\
B-Random Forest                                                                    & 88.8\%                                                    & 96.4\%                                                 & 92.5\%   & 89.0\%                                                                        & 96.4\%                                                                     & 92.5\%                      
\\ \hline
\end{tabular}
\end{table}

\end{document}